\documentclass[12pt]{article}
\usepackage{amssymb,amsmath}
\usepackage[noblocks]{authblk}
\usepackage[top=0.75in, bottom=0.75in, left=0.75in, right=0.75in, dvips]{geometry}
\usepackage{caption}
\begin{document}
\textwidth 10.0in 
\textheight 9.0in 
\topmargin -0.60in
\title{A Doubly Supersymmetric Particle in $3 + 3$ Dimensions}
\author[1,2]{D.G.C. McKeon}
\affil[1] {Department of Applied Mathematics, The
University of Western Ontario, London, ON N6A 5B7, Canada} 
\affil[2] {Department of Mathematics and
Computer Science, Algoma University, Sault St.Marie, ON P6A
2G4, Canada}
\date{}
\maketitle

\maketitle
\noindent
email: dgmckeo2@uwo.ca

\begin{abstract}
It is shown how in $3 + 3$ dimensions, it is possible to have a superparticle Lagrangian that has manifest supersymmetry both on the world line and in the target space.
\end{abstract}

\section{Introduction}
The so-called ``spinning'' particle [1,2] has a local manifest supersymmetry on the world line while the ``superparticle'' [3] has a local Fermionic symmetry in the target space [4] often called ``kappa symmetry''.  The nature of supersymmetry is affected by the properties of spinors in the space in which the supersymmetry is being considered [5,6].     It will be shown below that for the particular case of three space and three time dimensions, spinors have properties that make it relatively easy to obtain an action that is manifestly locally supersymmetric both on the world line and in the target space.  What makes it possible to be supersymmetric both in the target space and on the world line is that in $3 + 3$ dimensions one can have spinors $\kappa$ and $\rho$ that are both Majorana and Weyl, and satisfy the condition
\begin{equation}
\overline{\kappa} \,\tilde{\Gamma}^M \rho = - \kappa^T \tilde{\Gamma}^{MT} \overline{\rho}^T
\end{equation}
where $\tilde{\Gamma}^M$ is a Dirac matrix in $3 + 3$ dimensions.

The conventions used are in the appendix. 

\section{A Superparticle in $3 + 3$ Dimensions}

In order to have a Majorana-Weyl spinor satisfying eq. (1) and be in ten or less dimensions with at least three space and one time dimension, it is necessary to be in $3 + 3$ dimensions.  In no other dimension are all of these criterion met.

It follows from eq. (1) that if $\kappa$ and $\rho$ are Grassmann spinors (Fermionic) then
\begin{equation}
\overline{\kappa}\, \tilde{\Gamma}^M \rho = + \overline{\rho}\, \tilde{\Gamma}^M \kappa  
\end{equation}
while if either or both $\kappa$ and $\rho$ are ordinary spinors (Bosonic)
then 
\begin{equation}
\overline{\kappa}\,\tilde{\Gamma}^M \rho = - \overline{\rho}\,\tilde{\Gamma}^M \kappa \;.
\end{equation}

We now consider the spinning particle action [1,2]
\begin{align}
S &= \frac{1}{2} \int d\tau \left[ \eta_{\mu\nu} \left( \frac{\dot{\phi}^\mu (\tau) \dot{\phi}^\nu (\tau)}{e(\tau)} - i \psi^\mu (\tau) \dot{\psi}^\nu (\tau) \right.\right.\nonumber \\
&\left. \qquad - \frac{i}{e(\tau)} \chi(\tau) \dot{\phi}^\mu(\tau) \psi^\nu(\tau)\right] 
\end{align}
where $\eta_{\mu\nu}$ is the ``target space'' metric, the fields $\phi^\mu (\tau)$ (position) an $e(\tau)$ (einbein) are Bosonic and $\psi^\mu(\tau)$ (spin) and $\chi(\tau)$ (gravitino) are Fermionic (Grassmann).

By analysing the constraint structure of the action of eq. (3), it is possible to show that it is invariant under the transformations [7]
\begin{subequations}
\begin{align}
\delta \phi^\mu &= \frac{2B}{e} \left( \dot{\phi}^\mu - \frac{i}{2}\chi \psi^\mu\right) + i F\psi^\mu \\
\delta \psi^\mu &= \frac{F}{e} \left( \dot{\phi}^\mu - \frac{i}{2}\chi \psi^\mu\right)  \\
\delta e &= 2\dot{B} + iF\chi \\
\delta \chi &= 2\dot{F}\, .
\end{align}
\end{subequations}
In eq. (5), $B$ is a Bosonic gauge function and $F$ is a Fermionic gauge function.  The transformations of eq. (5) are not identical to these given in refs. [1,2].

In eq. (4) we now replace $\phi^\mu$ and $\psi^\mu$ in turn with 
\begin{subequations}
\begin{align}
\Phi^M &= \phi^M + \overline{\theta}\,\tilde{\Gamma}^M\theta \\
\Psi^M &= \psi^M + \overline{\lambda}\,\tilde{\Gamma}^M\theta 
\end{align}
\end{subequations}
where now we in $3 + 3$ dimensions (with indices $M$, $N \ldots$) and $\theta$ and $\lambda$ are Fermionic and Bosonic spinors respectively. Both are Majorana and Weyl.

On account of eqs. (2,3), the action of eq. (4) now becomes 
\begin{align}
S &= \frac{1}{2} \int d\tau\, \eta_{MN} \bigg[ \frac{1}{e} \left( \dot{\phi}^M + 2\overline{\theta} \,\tilde{\Gamma}^M \dot{\theta}\right) \left( \dot{\phi}^N + 2\overline{\theta} \,\tilde{\Gamma}^N \dot{\theta}\right) \nonumber \\
 & \hspace{1cm} -i \left( \psi^M + \overline{\lambda}\,\tilde{\Gamma}^M \theta\right) \left(\psi^N + \dot{\overline{\lambda}}\,\tilde{\Gamma}^N \theta - \overline{\theta}\tilde{\Gamma}^N \dot{\lambda}\right) \nonumber \\
 & \hspace{1cm} -\frac{i}{e} \chi \left(\dot{\phi}^M  + 2\overline{\theta} \,\tilde{\Gamma}^M \dot{\theta}\right)\left(\psi^N +  \overline{\lambda}\,\tilde{\Gamma}^N \theta \right)\bigg] \,.
\end{align}
From eqs. (2, 6a) we see that if
\begin{subequations}
\begin{align}
\delta_1\theta (\tau)& = \epsilon(\tau) \\
\intertext{\rm{and}}
\delta_1 \phi^M &= -2 \overline{\theta}\,\tilde{\Gamma}^M \epsilon = -2\overline{\epsilon} \, \tilde{\Gamma}^M \theta 
\intertext{\rm{where $\epsilon(\tau)$ is an infinitesimal Grassmann function of $\tau$ then}}
\delta_1 \Phi^M &= 0.
\end{align}
\end{subequations}
It also follows that
\begin{subequations}
\begin{align}
\delta_1 \Psi^M &= \delta \psi^M + (\delta \,\overline{\lambda}) \tilde{\Gamma}^M \theta + \overline{\lambda} \tilde{\Gamma}^M \epsilon 
\intertext{and so if}
\delta_1 \overline{\lambda} & = \chi \overline{\epsilon}\\
\delta_1 \psi^M & = - \chi \overline{\epsilon} \tilde{\Gamma}^M \theta - \overline{\lambda}\tilde{\Gamma}^M \epsilon\nonumber \\
&= \overline{\epsilon} \tilde{\Gamma}^M (\chi \theta + \lambda)
\intertext{then}
\delta_1\Psi^M = 0 \;.
\end{align}
\end{subequations}
Thus by eqs. (8c, 9d) we see that we have a local supersymmetry transformation in the target space that leaves the action of (7) invariant.

We now can substitute eq. (6) into the transformations of eqs. (5a,b) to obtain
\begin{subequations}
\begin{align}
\delta_2\left( \phi^M + \overline{\theta}\,\tilde{\Gamma}^M \theta \right) &= \frac{2B}{e}\left[ \left(\dot{\phi}^M + 2\overline{\theta}\,\tilde{\Gamma}^M\dot{\theta}\right) - \frac{i}{2}\chi \left( \psi^M + \overline{\lambda}\,\tilde{\Gamma}^M\theta\right)\right]\nonumber \\
& \hspace{1cm}+ iF\left( \psi^M + \overline{\lambda}\,\tilde{\Gamma}^M\theta\right)
\intertext{and}
\delta_2\left( \psi^M + \overline{\lambda}\,\tilde{\Gamma}^M \theta \right) &= \frac{F}{e}\left[ \left(\dot{\phi}^M + 2\overline{\theta}\,\tilde{\Gamma}^M\dot{\theta}\right) - \frac{i}{2}\chi \left( \psi^M + \overline{\lambda}\,\tilde{\Gamma}^M\theta\right)\right].
\end{align}
\end{subequations}
Again using eqs. (2,3), we see that if $\delta\phi^M$ is given by eq. (5a) and
\begin{equation}
\delta_2\theta = \frac{2B}{e} \left[ \dot{\theta} + \frac{i}{4} \chi \lambda \right] - \frac{iF}{2}\lambda 
\end{equation}
then eq. (10a) is satisfied.  Using eq. (11) we find that eq. (10b) becomes
\begin{align}
\delta_2 \psi^M + & \overline{\lambda}\,\tilde{\Gamma}^M \left[ \frac{2B}{e} \left(\dot{\theta} + \frac{i\chi}{4}\chi\lambda\right) - \frac{iF}{2}\lambda\right] - \overline{\theta}\,\tilde{\Gamma}^M (\delta\lambda) \\
&= \frac{F}{e}\left[\left( \dot{\phi}^M + 2\overline{\theta}\,\tilde{\Gamma}^M \dot{\theta}\right) - \frac{i\chi}{2} \left(\psi^M + \overline{\lambda} \,\tilde{\Gamma}^M \theta \right)\right]\, .\nonumber
\end{align}
Eq. (3) shows that $\overline{\lambda}\tilde{\Gamma}^M\lambda = 0$ and hence eq. (12) is satisfied provided
\begin{equation}
\delta_2\psi^M = \frac{F}{e} \left( \dot{\phi}^M - \frac{i}{2}\chi\psi^M \right) - \frac{2B}{e}\left(\overline{\lambda} \,\tilde{\Gamma}^M \dot\theta\right)
\end{equation}
and
\begin{equation}
\delta \lambda = \frac{F}{e} \left(2 \dot{\theta} - \frac{i}{2}\chi\lambda\right)\, .
\end{equation}
Together, eqs. (5a,c,d) and eqs. (11,13,14) constitute a set of Fermionic transformations on the world line that leave the action of eq. (4) invariant.

\section*{Discussion}

The action of eq. (4) has been shown to have a Fermionic gauge symmetry, both on the world line and in the target space, provided the target space has $3 + 3$ dimensions and that there we use Majorana-Weyl spinors.  Being in $3 + 3$ dimensions is crucial; for example the substitution of eq. (6a) would not work with Weyl spinors in $3 + 1$ dimensions as then $\overline{\theta}\,\tilde{\gamma}^\mu \theta$ automaticly vanishes.  However, it should be possible to compactify two of the temporal dimensions to reduce $3 + 3$ dimensions to $3 + 1$.

It would be of interest to quantize the model introduced here, or to devise an analogous string model.

\section*{Acknowledgements}
Roger Macleod had a helpful suggestion.

\section*{Appendix}
In six dimensional Euclidean space we use the Dirac matrices $\Gamma^M$
\[\Gamma^M = \left( \begin{array}{cc}
0 & i\gamma^\mu \\
-i\gamma^\mu & 0 \end{array}
\right)\;, \left( \begin{array}{cc}
0 & 1 \\
1 & 0 \end{array}
\right)\;; \Gamma^7 = i\Gamma^1 \ldots \Gamma^6 = \left( \begin{array}{cc}
-1 & 0 \\
0 & 1 \end{array}
\right)\eqno(A.1) \]
where $\gamma^\mu$ is a Dirac matrix in four Euclidean dimensions
\[\gamma^\mu = \left( \begin{array}{cc}
0 & i\sigma^i \\
-i\sigma^i & 0 \end{array}
\right)\quad \left( \begin{array}{cc}
0 & 1 \\
1 & 0 \end{array}
\right)\;; 
\gamma^5 = \gamma^1 \ldots \gamma^4 = \left( \begin{array}{cc}
-1 & 0 \\
0 & 1 \end{array}
\right)\; .\eqno(A.2) \]
(The Pauli spin matrices are $\sigma^i$.)  In moving to $3 + 3$ dimensional space, we define
\[ \tilde{\Gamma}^M = \Gamma^M \quad (M = 1 \ldots 3)\eqno(A.3)\]
\[\hspace{1cm} = i\Gamma^M \quad (M = 4 \ldots 6). \nonumber \]
If $\psi$ is a spinor in $3 + 3$ dimensions, we define
\[ \overline{\psi} = \psi^\dagger A^{-1} \eqno(A.4) \]
and
\[ \psi_C = C \overline{\psi}^T\, , \eqno(A.5) \]
where 
\[ A^{-1} \tilde{\Gamma}^M A = -\tilde{\Gamma}^{M\dagger} \eqno(A.6) \]
and
\[ C^{-1} \tilde{\Gamma}^M C = -\tilde{\Gamma}^{MT} \;.\eqno(A.7) \]
With the matrices $\Gamma^M$ given in eq. (A.1) it follows that
\[ A = \Gamma^4\Gamma^5\Gamma^6 = -A^{-1} = -A^\dagger = -A^T = A^\ast
\eqno(A.8) \]
and
\[ C = \Gamma^2\Gamma^4\Gamma^5 = -C^{-1} = -C^\dagger = -C^T = -C^\ast\;.
\eqno(A.9) \]
If $\psi$ is a spinor such that $\psi = \psi_C$ (Majorana condition) and $\psi = \Gamma^7 \psi$ (Weyl condition) then its eight components are of the form
\[ \psi = \left( 0,0,0,0,a,b, -ib^\ast , ia^\ast\right)^T \eqno(A.10) \]
where $a$ and $b$ are two independent complex numbers.  If $\rho$ and $\kappa$ are both Majorana and Weyl then eq. (1) follows.  Furthermore, we find that
\[ \overline{\rho}\kappa = 0 \eqno(A.11) \]
\[ \overline{\rho}\, \tilde{\Gamma}^M \tilde{\Gamma}^N \kappa = 0. \eqno(A.12) \]

\end{document}